\documentclass[twocolumn]{elsart}
\usepackage{epsfig}
\usepackage{graphics}
\usepackage[rflt]{floatflt}
\usepackage{tabularx}
\usepackage{amssymb}
\usepackage{amsfonts}
\usepackage{amsmath}
\usepackage{color}

\begin{document}

\begin{frontmatter}



\title{Beam Test Performance and Simulation of Prototypes for the ALICE Silicon Pixel Detector}

\author[label1]{J.~Conrad \corauthref{cor1}},
\author[label1]{G.~Anelli},
\author[label2]{F.~Antinori},
\author[label3]{A.~Badala},
\author[label3]{R.~Barbera},
\author[label1]{A.~Boccardi},
\author[label1]{M.~Burns},
\author[label4]{G.~E.~Bruno},
\author[label1,label4]{I.~A.~Cali},
\author[label1]{M.~Campbell},
\author[label4]{M.~Caselle},
\author[label1]{P.~Chochula},
\author[label1]{S.~Ceresa},
\author[label5]{M.~Cinausero},
\author[label2]{R.~Dima},
\author[label4]{D.~Elia},
\author[label2]{D.~Fabris},
\author[label5]{E.~Fioretto},
\author[label4]{R.~A.~Fini},
\author[label1]{S.~Kapusta},
\author[label1]{A.~Kluge},
\author[label6]{M.~Krivda},
\author[label4]{V.~Lenti},
\author[label3]{F.~Librizzi},
\author[label2]{M.~Lunardon},
\author[label4]{V.~Manzari},
\author[label1]{M.~Morel},
\author[label2]{S.~Moretto},
\author[label1]{A.~Morsch},
\author[label1]{P.~Nilsson},
\author[label1]{M.L.~Noriega},
\author[label1]{F.~Osmic},
\author[label3]{G.~S.~Pappalardo},
\author[label4]{V.~Paticchio},
\author[label2]{A.~Pepato},
\author[label3]{G.~Prete},
\author[label5]{A.~Pulvirenti},
\author[label1]{P.~Riedler}, 
\author[label3]{F.~Riggi},
\author[label6]{L.~Sandor},
\author[label4]{R.~Santoro},
\author[label2]{F.~Scarlassara},
\author[label7]{G.~Segato},
\author[label7]{F.~Soramel},
\author[label1]{G.~Stefanini},
\author[label1]{C.~Torcato de Matos},
\author[label2]{R.~Turrisi},
\author[label5]{L.~Vannucci},
\author[label2]{G.~Viesti},
\author[label8]{T.~Virgili}
\address[label1]{CERN, PH-EP Department, CH-1211 Geneva 23}
\address[label2]{Dipartimento di Fisica dell'Universit\`{a} and Sezione INFN, Padova, Italy}
\address[label3]{Dipartimento di Fisica dell'Universit\`{a} and Sezione INFN, Catania, Italy}
\address[label4]{Dipartimento di Fisica dell'Universit\`{a} and Sezione INFN, Bari, Italy}
\address[label5]{Laboratori Nazionali di Legnaro, Legnaro, Italy}
\address[label6]{Institute of Experimental Physics, Slovak Academy of Sciences, Kosi\v{c}e,Slovakia}
\address[label7]{Dipartimento di Fisica dell'Universit\`{a} and Sezione INFN Gruppo Collegato di Udine, Udine, Italy}
\address[label8]{Dipartimento di Fisica dell'Universit\`{a} and Sezione INFN, Salerno, Italy}

\corauth[cor1]{Corresponding author: Jan Conrad, PH-EP Dept, F01910, CH-1211 Geneva 23}

\noindent

\begin{abstract}
The silicon pixel detector (SPD) of the ALICE experiment in preparation at the Large Hadron Collider (LHC) at CERN is designed to provide the precise vertex reconstruction needed for measuring heavy flavor production in heavy ion collisions at very high energies and high multiplicity.
The SPD forms the innermost part of the Inner Tracking System (ITS) which also includes silicon drift and silicon strip detectors. Single assembly prototypes of the ALICE SPD have been tested at the CERN SPS using high energy proton/pion beams in 2002 and 2003. We report on the experimental determination of the spatial precision. We also report on the first combined beam test with prototypes of the other ITS silicon detector technologies at the CERN SPS in November 2004. The issue of SPD simulation is briefly discussed.
\end{abstract}

\end{frontmatter}


\newpage

\section{Introduction}
\begin{figure} [htb]
\begin{center}
\includegraphics*[width=.2\textwidth]{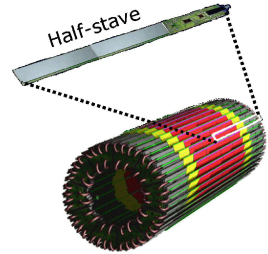}
\end{center}
\caption{The SPD barrel and a half stave.}
\label{fig:barrel}
\end{figure}

The ALICE experiment at the LHC is 
primarily dedicated to study heavy ion collisions. Such collisions produce a high multiplicity environment (up to 8000 tracks per unit rapidity) which requires excellent tracking capabilities, in particular for secondary vertex reconstruction for the study of beauty and charm physics. The detector system dedicated to this task is the ITS \cite{its_tdr}. The SPD (see figure \ref{fig:barrel}) constitutes the two innermost layers (situated at radii 3.9 and 7.2 cm, respectively) of the ITS, which also includes two layers of silicon drift and two layers of silicon strip detectors at larger radii. Each SPD module (``ladder'') consists of five readout chips bump bonded to a p-in-n sensor of
$200~{\mu m}$ thickness. Each chip contains 8192 read-out cells corresponding to pixels of dimension $50~{\mu m}$ ($r\phi$) $\times$ $425~{\mu m}$ (z). Two ladders together with on-detector read out electronics (Multi-Chip Module) form a half-stave. Two half-staves are combined to cover the whole length of the SPD barrel (28.6 cm). The total number of staves in the SPD is 60 (20 in the inner layer, 40 in the outer), with in total 1,200 read-out chips and 9.83 million read-out channels. The acceptance of the SPD is $|\eta| \sim 1.8$ and $|\eta| \sim 0.8$ for the inner and outer layer, respectively.
\section{Beam Tests at CERN SPS}
\begin{figure} [htb]
\begin{center}
\includegraphics*[width=.4\textwidth]{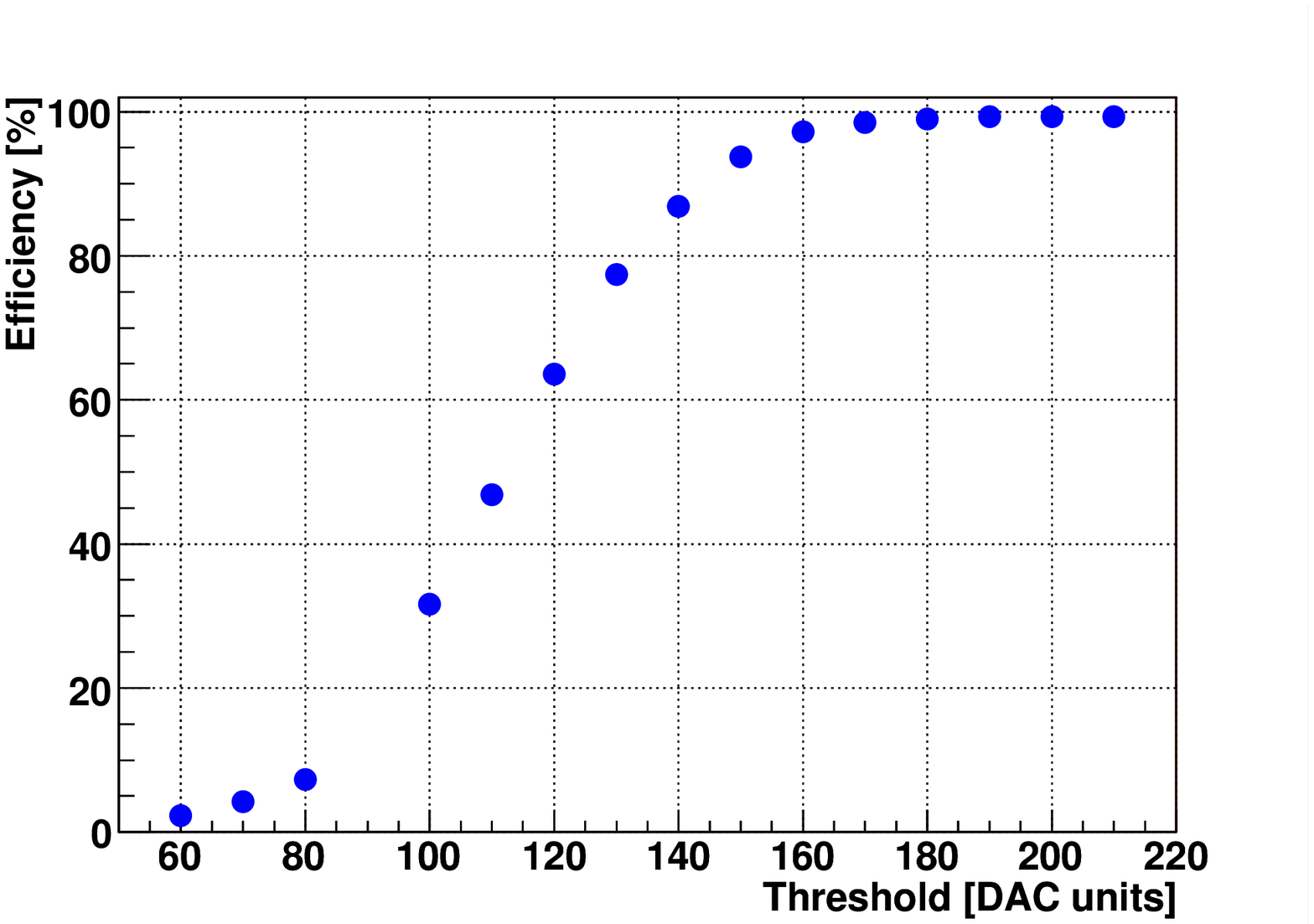}
\end{center}
\caption{The reconstruction efficiency as a function of threshold for $200~{\mu m}$ thick sensors. The threshold shown is the setting of a 8-bit DAC such that a lower DAC setting corresponds a to higher threshold. A setting of 214 is equivalent to approximately 2000 e$^-$. The normal working point is around DAC = 200.}
\label{fig:efficiency}
\end{figure}

Prototypes of the SPD modules in the form of single assemblies in a stand-alone configuration, have been tested in high energy particle beams at the CERN SPS. The primary objective was the validation of the pixel ASICs and of the read-out electronics. The setup was not optimized for precision measurements of the position resolution, but an accurate determination of the intrinsic spatial precision could anyway be performed. Data runs were taken  in 2002 and 2003 with proton/piolsn beams of 350 GeV/c and 120 GeV/c momentum, respectively.  The set-up consisted of a similar basic configuration: 4 reference planes were used to define tracks which were then projected into a test plane which was situated in between the reference planes. In the 2003 run, a prototype full read-out chain with two ladders was installed as part of the configuration and read out separately. Many parameters of the pixel ASIC, in particular the threshold, can be remotely adjusted via on-chip digital to analog converters (DAC). The prototype in the test plane was studied under different conditions (threshold scan, different inclination angles w.r.t. the beam and bias voltage scan). Clusters of hit pixels correlated with a track were then used to estimate the combined detector/reconstruction efficiency, which was found to be $>$ 99 \% in a wide range of threshold values including the normal working point, see figure \ref{fig:efficiency}.

\begin{figure} [htb]
\begin{center}
\includegraphics*[width=.4\textwidth]{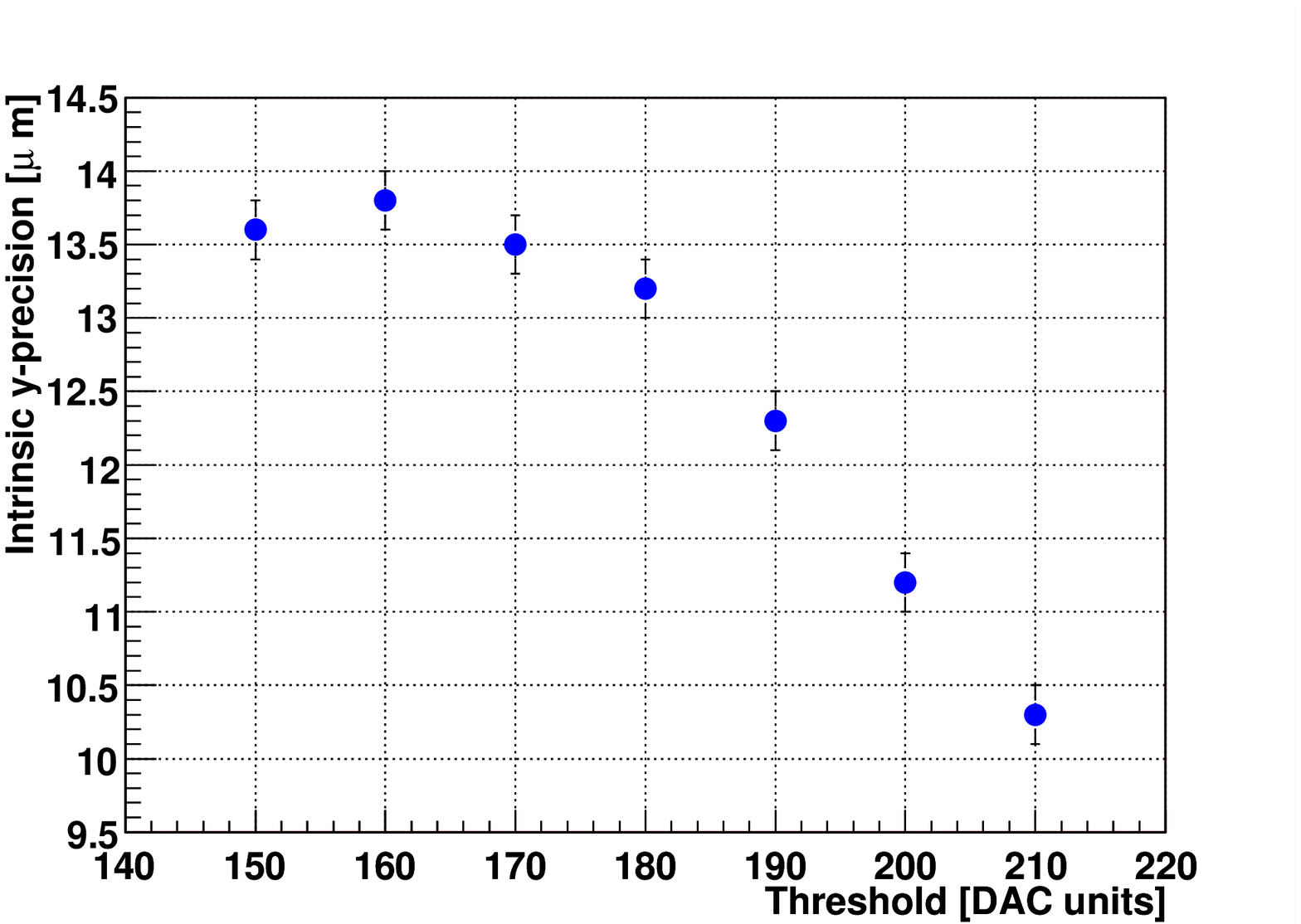}
\includegraphics*[width=.4\textwidth]{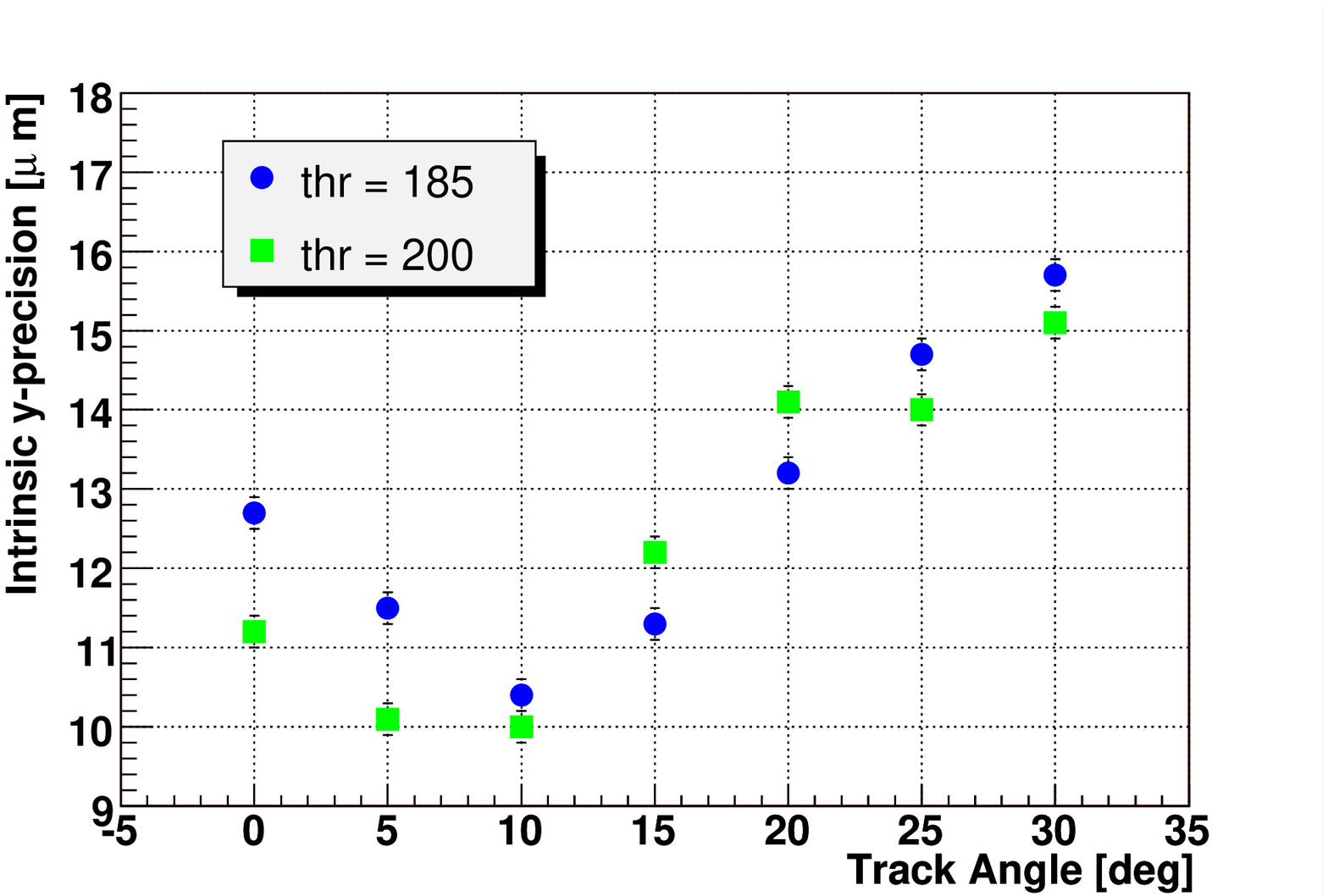}

\end{center}
\caption{The intrinsic precision (200  ${\mu m}$ sensor) as a function of threshold for tracks normal incidence angle (upper figure) and as a function of angle for two different thresholds (lower figure)}
\label{fig:resolution}
\end{figure}

The intrinsic precision as a function of various parameters has been calculated using an iterative method \cite{Conrad::2005}\cite{Elia:2005}.
The intrinsic precision at the normal working point is found to be ( 11.1 $\pm$ 0.2 ) ${\mu m}$ in the $r\phi$ direction.
The dependence of intrinsic precision on threshold and angle of incidence is shown in figure \ref{fig:resolution}. 
A higher multiplicity region was also investigated using a 158 A GeV/c Indium ion beam on a Pb target. The data analysis is under way.
Discussions of the 2002 and 2003 SPD beam tests can also be found in \cite{Elia:2005} \cite{Nilsson:2004kf} and \cite{Riedler::2002}.\\
\noindent
In 2004 a combined ITS beam test was carried out. The setup consisted of two planes of SPD (each equipped with one half stave), two planes of SDD and two planes of SSD. For the SPD, the full read-out chain was tested using the final components. As part of the integration test, the ALICE DAQ and trigger system were for the first time used for more than one sub-detector system. Furthermore, beam test data is used for code validation of the ALICE software framework AliRoot \cite{AliRoot}.
\section{Simulation of the Silicon Pixel Detector}
\begin{figure} [htb]
\begin{center}
\includegraphics*[width=.4\textwidth]{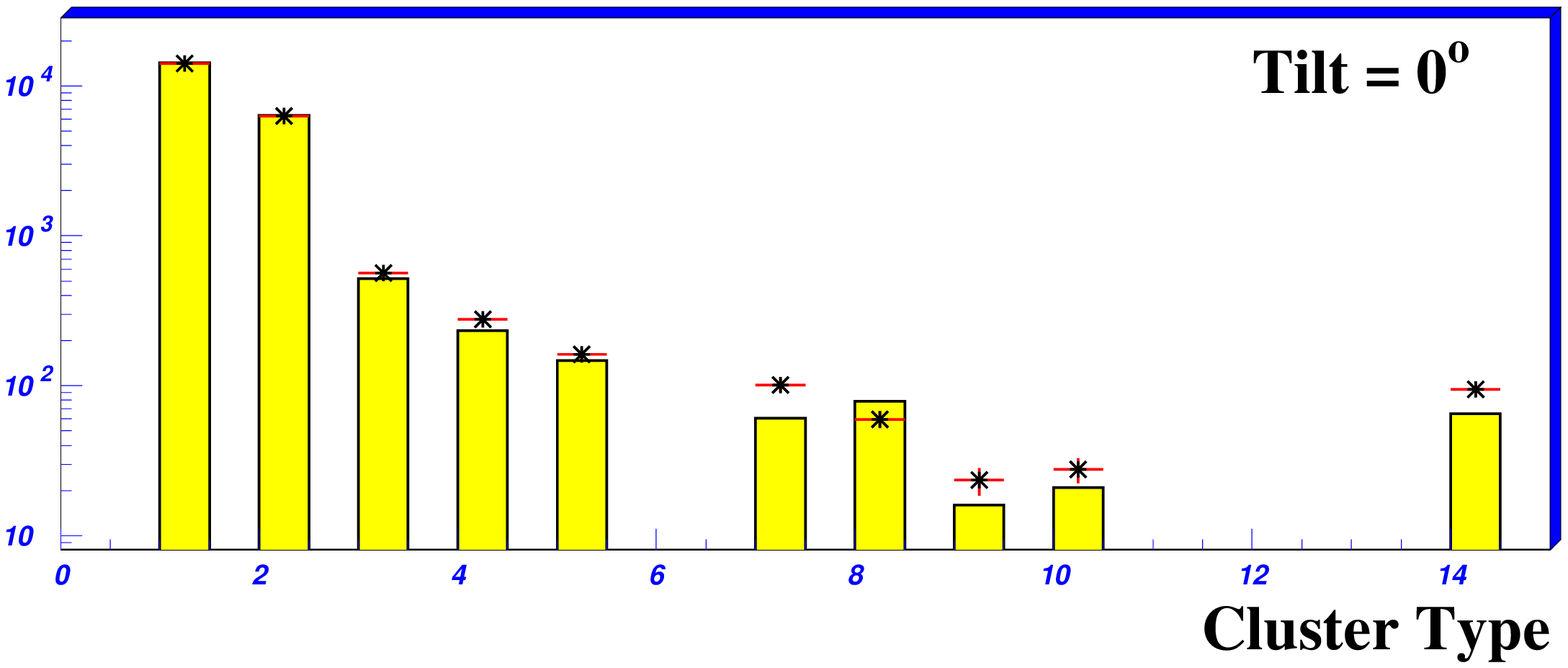}
\includegraphics*[width=.4\textwidth]{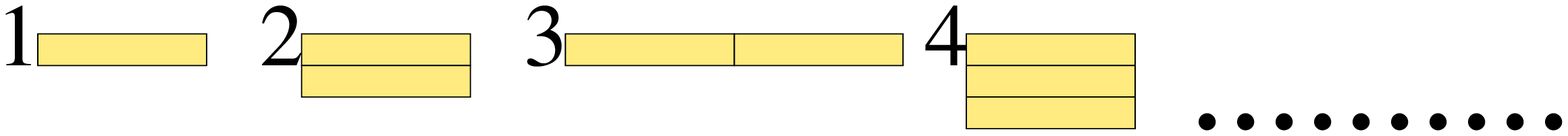}
\end{center}
\caption{The distribution of different cluster types for simulation (histogram) and from data from the 2002 beam test (stars). The definition of the four most common cluster types is also given. For the definition of the remaining cluster types, see \cite{Elia:2005}.}
\label{fig:clustertype}
\end{figure}

Particle tracking in ALICE is currently based on GEANT3 {\cite{Brun}}. The energy deposited by GEANT is transformed into the number of electron-hole pairs (3.6 eV/(e,h)) and then compared with a threshold. Threshold fluctuations and noise are taken into account pixel by pixel. Charge sharing is simulated assuming Gaussian diffusion of e/h-pairs. In each GEANT step, the diffusion variance is evaluated: $\sigma_{\text{diff}} = k\sqrt{l_{\text{dr}}}$, where $l_{\text{dr}}$ is the drift path and $k =\sqrt{2D/v_{\text{dr}}}$. $D$ is taken to be the hole diffusion coefficient: $D$ = 11 cm $^2/$s \cite{Sailor:1991ky} and $v_{\text{dr}} = \mu \cdot E$, where the hole mobility is taken to be $\mu$ = 450 cm$^2/$Vs \cite{Sailor:1991ky} and $E$ is the electric field per unit length (which depends on the bias voltage). This model gives a qualitatively good description of the beam test data (see figure \ref{fig:clustertype}) and tuning of the model parameters is currently under way.\\
GEANT3 is not supported anymore since 1996. Therefore, ALICE implements the possibility to use GEANT4 \cite{Agostinelli:2002hh} and the more broadly physics validated FLUKA \cite{Fasso:2000hd} in its detector simulation. The code which interfaces the ALICE software package AliRoot \cite{AliRoot} with these packages has been validated using the beam test of 2004. For example, it is confirmed that differences between FLUKA and GEANT3 do not affect the simulation of the SPD. 
\section{Conclusions}
Data analysis of the proton beam tests in 2002 and 2003 shows excellent performance of the ALICE SPD with $>$ 99 \% efficiencies and intrinsic precision $\sim$ 11 $\mu m$ in the $r\phi$ coordinate. Given the successful exploitation of the previous beam tests, the 2004 beam test integrating all three ITS silicon detector technologies did not need to entirely focus on detector and read out electronics performance. The detectors were rather used to test ALICE DAQ, trigger and data analysis which is used to validate the ALICE software framework.\\Simulation of the SPD using GEANT3 including Gaussian diffusion of charges gives a good description of the beam test data of 2002 and 2003 already with simple assumptions on the model parameters. Further improvement is expected from currently performed tuning of its parameters. Newer particle tracking packages, like GEANT4 or the more extensively physics validated FLUKA are implemented within the ALICE detector simulation and the code is validated using the 2004 beam test data.

\end{document}